%% file: main.tex
\documentclass[runningheads]{llncs}
\usepackage{booktabs}
\usepackage{array}
\usepackage{multirow}
\usepackage{geometry}
\usepackage{listings}
\usepackage[T1]{fontenc}
\usepackage{graphicx}
\usepackage[hyphens]{url}
\begin{document}
\title{Cyber Warfare During Operation Sindoor: Malware Campaign Analysis and Detection Framework}
 \author{Prakhar Paliwal %\inst{1}\orcidID{0000-1111-2222-3333}
 \and
Atul Kabra 
%\inst{2,3}\orcidID{1111-2222-3333-4444} 
\and
Manjesh Kumar Hanawal
%\inst{3}\orcidID{2222--3333-4444-5555}
}
% %
 \authorrunning{ Prakhar Paliwal et al.}
 \institute{Indian Institute of Technology Bombay, Mumbai, Maharashtra
 %\and
\\ \email{\{prakhar.paliwal, 25D2020, mhanawal\}@iitb.ac.in}
}
% %
\maketitle              % typeset the header of the contribution
\begin{abstract}
Rapid digitization of critical infrastructure has made cyberwarfare one of the important dimensions of modern conflicts. Attacking the critical infrastructure is an attractive pre-emptive proposition for adversaries as it can be done remotely without crossing borders. Such attacks disturb the support systems of the opponents to launch any offensive activities, crippling their fighting capabilities. Cyberattacks during cyberwarfare can not only be used to steal information, but also to spread disinformation to bring down the morale of the opponents. Recent wars in Europe, Africa, and Asia have demonstrated the scale and sophistication that the warring nations have deployed to take the early upper hand. In this work, we focus on the military action launched by India, code-named Operation Sindoor, to dismantle terror infrastructure emanating from Pakistan and the cyberattacks launched by Pakistan. In particular, we study the malware used by Pakistan APT groups to deploy Remote Access Trojans in Indian systems. We provide details of the tactics and techniques used in the RAT deployment and develop a telemetry framework to collect necessary event logs using Osquery with a custom extension. Finally, we develop a detection rule that can be readily deployed to detect the presence of the RAT or any exploitation performed by the malware.

\keywords{Operation Sindoor, Malware Campaign Analysis, Osquery, Malware detection, APTs, Cyberwarfare.}
\end{abstract}

\section{Introduction}
\label{sec:Intro}
\input{Intro}

\section{Operation Sindoor Campaign}
\label{sec:OPSindoor}
\input{OPSindoor}

\section{Logging events and Visibility}
\label{sec:Logging}
\input{Logging}

\section{Analysis and Detections}
\label{sec:Telemetry}
\input{Telemetry}

\section{Conlusions}
\label{sec:Conclusion}
\input{Conclusion}

\bibliographystyle{splncs04_unsrt}
\bibliography{Biblio.bib}

% % \input{Biblio.tex}
% \end{thebibliography}

\end{document}

%% file: Intro.tex
% Cyber attacks are part of the modern warfare that get activated at the onset of the conflicts before any firepowers are launched. Recent conflicts between Russia and Ukraine, Iran and Israeal, Indian and Pakistan, Cambodia and Thailand led to several cyber attacks on their opponents. In this work we will focus on the cyberattacks on India during the Operation Sindoor. 

% Operation Sindoor was commenced by Indian security forces as a tactical response to the terrorist assault in Pahalgam in April 2025, which distinctly targeted non-combatants. This operation, noted for its precise and focused military interventions against terrorist infrastructure, was executed without violating international boundaries and signified a substantial advancement in India's strategic defense posture. 

Cyberattacks have become an integral component of modern warfare. They are activated at the onset of the conflicts even before any firepowers are launched or when the hostilities are imminent. Critical infrastructure like power, telecom networks, nuclear power plants, and financial systems that are heavily reliant on cyber space, become the primary target of attackers. In addition to crippling the infrastructure, attackers also resort to stealing information, spreading disinformation, and deface government portal to bring down the morale of its opponents. Recent conflicts between Russia and Ukraine \cite{Russia-Ukraine}, Iran and Isreal \cite{Iran-Israel}, Indian and Pakistan \cite{India-Pakistan}, Cambodia and Thailand  \cite{Thailand-Cambodia} led to several cyber attacks between the warring nations. These examples reveal that cyber operations are now woven into the fabric of modern warfare, shaping both the course and outcome of physical confrontations. In this work we focus on military operation, code named `Operation Sindoor' launched by India and the cyberattack it faced. We will discuss the modus-operandi of the attackers and develop a framework to detect the attack.

%Cyber attacks from state sponsored groups are the first sign of escalting tension between countries as has been seen in recent history events such as 

Operation Sindoor, launched in May 2025 by Indian military forces as a tactical response to the terrorist assault in Pahalgam on 22nd April where attackers shot several unarmed male tourists at point-blank range in front of their pleading families. Operation Sindoor was characterized by precise and focused strikes against terrorist infrastructure, executed without violating international borders — a reflection of India’s growing emphasis on strategic, measured defense postures \cite{PIB}.  While the physical military operations unfolded at borders with firepower crossing borders, a quieter and less visible battle was waged in the cyberspace. Indian networks across government departments, defense institutions, and critical civilian sectors began to experience a wave of cyber attacks, suggesting that the punitive action initiated by Indian on the terror infrastructure had triggered a coordinated offensive cyber campaigns by Pakistani groups.

While the primary focus of offensive cyber campaigns is to sabotage critical infrastructure needed to support military operation and deprive a nation from using its full military capabilities, it also aims to manipulate information flows to bias public opinions. Parallelly the attackers can operate cyber espionage in the shadows, quietly penetrating networks to gather political, military, or economic intelligence over extended periods, often without immediate visible disruption. All forms were likely at play during Operation Sindoor, as adversaries sought not only to weaken India’s operational capabilities but also steal sensitive data and manipulate public opinion. %In the aftermath of the Pahalgam assault, these tactics offered terrorist groups an additional vector to retaliate and influence public perception.

%Layered onto these are the more overt and psychologically driven tactics of cyber terrorism, in which non-state actors employ digital means to cause fear, disrupt essential services, or project power far beyond their physical reach. Whether through denial-of-service attacks on government portals, defacement of official websites, or the spread of targeted propaganda, such actions are intended to create uncertainty and amplify the perceived reach of the attacker. 
%In the aftermath of the Pahalgam assault, these tactics offered terrorist groups an additional vector to retaliate and influence public perception.

Behind these categories of cyber activity lies a diverse ecosystem of actors, each playing distinct yet often interconnected roles. Advanced Persistent Threat (APT) groups — elite, well-resourced groups frequently tied to state intelligence agencies — operate as the long-term specialists of cyber conflict, infiltrating networks with the patience to remain undetected for months or even years. Their operations are strategic, designed to align with national objectives, whether that means pre-positioning for sabotage or extracting intelligence critical to military planning. Alongside them, more transactional entities such as hack-for-hire groups and cyber mercenaries provide on-demand capabilities, allowing states and other clients to commission specific operations without direct involvement. These actors offer both expertise and plausible deniability, enabling campaigns that are precise, targeted, and deniable.

During the Operation Sindoor, a malware was propagated through spear phishing emails that targeted government officials. The malware was sent as an email attachment disguised as a detailed report on Operation Sindoor. The report included a Remote Access Trojan (RAT) that would trigger when a user opened the attached file. The RAT then connected to a command and control server enabling the attackers to execute more than 20 commands on the endpoint remotely. The commands enabled the remote attackers to perform various activities like screen capture, delete file, and exfiltrate files to C2. 
This in turn could serve as a pipeline to offload other stronger malware or ransomware to exfiltrate and encrypt data.

Attackers can deploy a combination of tactics, techniques and procedures (TTPs) as listed in the Mitre Attack framework~\cite{mitre_attack}. For example, they can use various techniques like spearphishing to obtain initial access to the victim machine, and then use the payload (RAT) to execute malicious commands. Having visibility to events corresponding to these TTPs is important to detect the attacks. The behaviour of the malware during Operation Sindoor are well documented in \cite{seqrite_1,seqrite_2}. 

Several tools are available to collect information about activities performed by malware by logging  system activities on the endpoints. We identify all the necessary event that needs to be monitored to detect the malware. We use Osquery \cite{osquery}\cite{osquery_1}, an endpoint operating system orchestration tool, to gather all the relevant activities of the malware . The default windows version of Osquery does not support collection of all the relevant logs. But thanks to the extensibility feature of Osquery, we use a custom extension to enhance the default capability of Osquery to gather additional logs related to the activities performed by the malware.
%We use custom extensions to gather full visbility of the chain of activities done by the malware. 

We map the events in the logs to the activities of the malware and establish correlation across various events like process spawning, file and network activities to develop a detect if an endpoint is subject to the malware attack. Also, our framework allows to detect presence of the malware in early stages and prevent it from performing any malicious activities. Finally, we develop a SQL query that can be run on an endpoint with Osquery with the custom extension to detect the presence of the malware. In summary, our contributions are as follows:
\begin{itemize}
    \item Identify the events that needs to be logged for telemetry
    \item Develop a framework to detect the malicious behaviour from the telemetry 
    \item An SQL rule to detect if an endpoint is subject to the malware attack. 
\end{itemize}
The rest of the paper is organized as follows: In Section \ref{sec:OPSindoor}, we provide details about the malware used in the cyber campaign during Operation Sindoor. In Section \ref{sec:Logging}, we discuss about the logging the necessary events to establish telemetry. In Section \ref{sec:Telemetry}, we provide details about analysis of the telemetry logs and develop a rule to detect the malware. We end with conclusions in Section \ref{sec:Conclusion}.

\subsection{Related Works}
% Cybersecurity research has long examined the evolving landscape of attack campaigns, detection mechanisms, and incident analysis frameworks, providing valuable foundations for understanding large-scale and politically motivated cyber operations. Such operations, often blending cyber attacks or digital incursions with broader strategic objectives, have been documented in diverse contexts, from state-sponsored cyber espionage to coordinated cyber warfare by hactivist groups, APTs or just `hack for hire' groups. Below we discuss papers that analyse various malware campaigns during cyber warfare. 

Cybersecurity research has long explored attack campaigns, detection mechanism, and incident analysis frameworks, offering insights into large-scale and politically motivated cyber operations. 
A survey on cyber warfare analyzes the domain across its physical, syntactic, and semantic layers, outlining common attack vectors such as reconnaissance, access, denial-of-service, and espionage, and highlighting the motivations, tools, and potential impacts ranging from disruption of essential services to national security threats\cite{cyber_Survey}. Such operations, from state-sponsored espionage to hacktivist or “hack-for-hire” campaigns, often start with reconnaissance. Any operation, spanning from state-sponsored espionage to hacktivist or a “hack-for-hire” campaigns, often begin with reconnaissance. 
A survey on cyber scanning which is often the first step of any kind of attack shows how it helps attackers find targets, spot weaknesses, and plan large-scale, coordinated, and stealthy attacks\cite{cyber_scanning}.
Below, we review studies on malware campaigns in cyber warfare.

{\bf SSH attacks:} One major area of investigation has focused on the persistence and tactics of intruders during prolonged intrusion attempts. \cite{khandait2021} analyzes SSH brute-force activity in a production environment, revealing patterns of repeated credential guessing, attacker geolocation, and the recurrence of specific adversary infrastructure over weeks. Their use of flow-based network features with machine learning classifiers achieved high accuracy in differentiating successful from failed logins, demonstrating the feasibility of scalable intrusion detection in real-world campaigns.

{\bf Visibility and telemetry:} Advanced endpoint detection and response (EDR) solutions, such as those studied by Medisetti et al. \cite{prasad2024}, integrate behavioral analytics, host telemetry, and historical baselines to detect anomalies indicative of lateral movement and data exfiltration. Such endpoint-level intelligence complements network telemetry, offering visibility into the deeper stages of an attack lifecycle.
Also, just looking into the telemetry with the use of native windows tool like sysmon and ETW (Event Trace for Windows) may not be enough as they are not 'evented' \footnote{Logs are collected not just when the query is launched, but all between the queries so that no log is lost in-between the queries} . Current cybersecurity related incidents requires telemetry which can be logged asynchronously over time in an evented manner to perform correlation analysis \cite{vajra1}. The native Auditd demon in Linux offers contextualized log collections, but has problems with log forwarding, storing and parsing for future use \cite{vajra2}.

{\bf Cyber terrorism:} The strategic dimension of cyber operations has also been examined in the context of cyber terrorism and information warfare. \cite{mosoiu2020} documented how politically motivated actors combine cyber intrusions with psychological and information operations to disrupt governance and destabilize societies. Aggrey et al. \cite{aggreyanalysing} extended this by analyzing recent advanced persistent threat (APT) incidents, illustrating how attackers align technical compromises with long-term strategic objectives. In addition, O’Brien \cite{obrien2022} emphasized the role of modern security frameworks and threat intelligence integration in countering advanced state-sponsored actors, such as the Russian group Cozy Bear aka APT-29. These insights are directly relevant to Operation Sindoor, where attribution, capability assessment, and defensive readiness form an integral part of campaign analysis done by APT groups.

{\bf Regional Campaign analysis:} Regional threat intelligence further enriches campaign analysis by situating incidents within their geopolitical context. \cite{sarowa2022} examined cyberattack patterns across APCERT member nations, identifying prevalent malware families, favored attack vectors, and cross-border incident trends. These findings underscore the importance of contextualizing Operation Sindoor within the broader Asia-Pacific threat environment, where geopolitical tensions and regional alliances can influence both targeting and attacker behavior.

While such detection-oriented studies capture the operational signatures of attackers, other works provide structured frameworks for understanding their tactics and planning effective defences. \cite{stango2009} proposed a threat analysis methodology that maps adversary actions to the cyber kill chain, enabling security teams to evaluate vulnerabilities, anticipate attack stages, and prioritize countermeasures. This systematic approach is particularly applicable when examining campaigns that unfold in multiple coordinated phases, as in the case of Operation Sindoor.

Attacks are not limited to endpoints or traditional IT systems any more. Cyber-physical power systems have been targeted by attacks that exploit weaknesses in the physical, cyber, and control layers, and researchers have reviewed these attack types along with detection and defense methods, from state-estimation approaches to machine learning-based solutions \cite{cyber_power}, In a similar manner, IIoT systems are also being targeted, as they face attacks such as denial-of-service, data breaches, and malware injection, targeting different layers of the architecture and there is a need to develop effective defences which increases the security of IIoT systems \cite{cyber_iiot}.

Taken together, these studies paint a comprehensive picture of how cyber campaigns are planned, executed, and detected. They also reveal the necessity of combining multiple analytical perspectives — from flow-based detection and endpoint analytics to geopolitical context and strategic threat modelling — in order to fully understand and counter complex, targeted operations. This multi-layered approach forms the basis for the methodology applied in the present study of the Operation Sindoor cyberattack campaign.

%% file: OPSindoor.tex
%Operation Sindoor was commenced by Indian security forces as a tactical response to the terrorist assault in Pahalgam in April 2025, which distinctly targeted non-combatants. This operation, noted for its precise and focused military interventions against terrorist infrastructure, was executed without violating international boundaries and signified a substantial advancement in India's strategic defense posture. %The operation illustrated the profound integration of indigenous technological innovations such as unmanned aerial vehicle operations, multi-layered air defense mechanisms, and electronic warfare systems.

%As in modern warfare, Operation Sindoor also escalated into cyber warfare. Malicious actors, capitalizing on the escalated tensions subsequent to the Pahalgam incident, launched intricate cyber espionage initiatives aimed at infiltrating sensitive networks of the Indian government. These assaults predominantly utilized targeted spear-phishing communications, meticulously designed to appear legitimate, exploiting the emotional urgency linked to recent geopolitical occurrences. The emails contained attachments, including macro-enabled documents and deceptive PDF files, PPT files and html files which, upon interaction, dropped and executed remote access trojans (RATs), specifically Crimson RAT and associated malware frameworks.

Operation Sindoor, originally a military response to the terrorist attack in Pahalgam on April 22, 2025, swiftly evolved into a sophisticated cyber campaign orchestrated by  Advanced Persistent Threat (APT) group APT36, also known as Transparent Tribe\cite{APT36}, capitalizing on the incident’s geopolitical significance. Telemetry data, identified anomalous spear-phishing traffic targeting Indian governmental and defence networks commencing April 17, 2025, five days prior to the attack, suggesting preemptive reconnaissance activities\cite{seqrite_1}. By April 24, 2025, malicious documents exploiting the Pahalgam attack, such as “Action Points \& Response by Govt Regarding Pahalgam Terror Attack.pdf” (authored under the pseudonym “Kalu Badshah”), proliferated across public domains, hosted on fraudulent websites including \textit{jkpolice[.]gov[.]in[.]kashmirattack[.]exposed} and \textit{pahalgamattack[.]com} \cite{seqrite_1}. These domains, registered within 48 hours post-attack and hosted across autonomous systems such as AS 200019 (Alexhost Srl) and AS 213373 (IP Connect Inc), impersonated reputable Indian entities, notably the Jammu \& Kashmir Police and Indian Air Force, to facilitate credential harvesting and surreptitious data exfiltration. Such cyber intrusions underscored the vulnerabilities in existing cybersecurity infrastructure, highlighting the critical need for enhanced cybersecurity readiness, training, and resilient defensive frameworks.The strategic ramifications of these cyber assaults during Operation Sindoor included compromised communications, and an  elevated risks of sensitive data breaches, thereby amplifying geopolitical tensions, highlighting the campaign’s role in contemporary hybrid warfare dynamics.

Coordinated cyber incidents targeting critical sectors, including defense, government IT, healthcare, telecommunications, and education were recorded during the period when Operation Sindoor was taking place\cite{seqrite_2}. 

Spear-phishing emails distributed weaponized files, masquerading as official advisories to exploit heightened public concern over national security. These emails featured a diverse array of malicious file formats—PDF, PPTX.LNK, PPAM, XLAM, XLSB, and MSI—to circumvent security protocols and exploit user trust. For instance, the PDF document \textit{“Action Points \& Response by Govt Regarding Pahalgam Terror Attack.pdf”} embedded hyperlinks redirecting to fraudulent login pages designed for credential phishing, were hosted on domains like \textit{jkpolice[.]gov[.]in[.]kashmirattack[.]exposed}. A full list of file names and malicious domains can be found in Table \ref{tab:docs} and \ref{tab:domains} respectively.

\begin{table}
\centering
\caption{Compilation of Documents}
\label{tab:docs}
\begin{tabular}{|l|l|l|}
\hline
\textbf{S.\,No.} & \textbf{Document} & \textbf{Format}\\
\hline
1  & Report \& Update Regarding Pahalgam Terror Attack                       & PDF\\
2  & Report Update Regarding Pahalgam Terror Attack                          & PDF\\
3  & Action Points \& Response by Govt Regarding Pahalgam Terror Attack      & PDF\\
4  & J\&K Police Letter                                                      & PDF\\
5  & ROD on Review Meeting held on 10 Apr 2025 by Secy DRDO                  & PDF\\
6  & Record of Discussion – Technical Review Meeting Notice                  & PDF\\
7  & Meeting Notice – 13\textsuperscript{th} JWG meeting (India – Nepal)      & PDF\\
8  & Agenda Points for Joint Venture Meeting at IHQ MoD                      & PDF\\
9  & DO Letter, Integrated HQ of MoD                                         & PDF\\
10 & Collegiate Meeting Notice \& Action Points – MoD                        & PDF\\
11 & Letter to the Raksha Mantri Office                                      & PDF\\
12 & (Unnamed file “pdf”)                                                    & PDF\\
13 & Alleged Case of Sexual Harassment by Senior Army Officer                & PDF\\
14 & Agenda Points of Meeting of Dept of Defence                             & HTML\\
15 & Action Points of Meeting of Dept of Defence                             & HTML\\
16 & Agenda Points of Meeting of External Affairs Dept                       & HTML\\
\hline
\end{tabular}
\end{table}

\begin{table}
\centering
\caption{Phishing Domains and Associated IP Addresses}
\label{tab:domains}
\begin{tabular}{|l|l|l|}
\hline
\textbf{S.\,No.} & \textbf{Domain Name} & \textbf{IP Address(es)}\\
\hline
1  & jkpolice[.]gov[.]in[.]kashmirattack[.]exposed          & 37.221.64.134,\;78.40.143.189\\ \hline
2  & iaf[.]nic[.]in[.]ministryofdefenceindia[.]org          & 37.221.64.134\\ \hline
3  & email[.]gov[.]in[.]ministryofdefenceindia[.]org        & 45.141.58.224\\ \hline
4  & email[.]gov[.]in[.]departmentofdefenceindia[.]link     & 45.141.59.167\\ \hline
5  & email[.]gov[.]in[.]departmentofdefence[.]de            & 45.141.58.224\\ \hline
6  & email[.]gov[.]in[.]briefcases[.]email                  & 45.141.58.224,\;78.40.143.98\\ \hline
7  & email[.]gov[.]in[.]modindia[.]link                     & 84.54.51.12\\ \hline
8  & email[.]gov[.]in[.]defenceindia[.]ltd                  & 45.141.58.224,\;45.141.58.33\\ \hline
9  & email[.]gov[.]in[.]indiadefencedepartment[.]link       & 45.141.59.167\\ \hline
10 & email[.]gov[.]in[.]departmentofspace[.]info            & 45.141.58.224\\ \hline
11 & email[.]gov[.]in[.]indiangov[.]download                & 45.141.58.33,\;78.40.143.98\\ \hline
12 & indianarmy[.]nic[.]in[.]departmentofdefence[.]de       & 176.65.143.215\\ \hline
13 & indianarmy[.]nic[.]in[.]ministryofdefenceindia[.]org    & 176.65.143.215\\ \hline
14 & email[.]gov[.]in[.]indiandefence[.]work                & 45.141.59.72\\ \hline
15 & email[.]gov[.]in[.]drdosurvey[.]info                   & 192.64.118.76\\
\hline
\end{tabular}
\end{table}

\begin{figure}[h]
    \centering
    \includegraphics[width=1.0\textwidth]{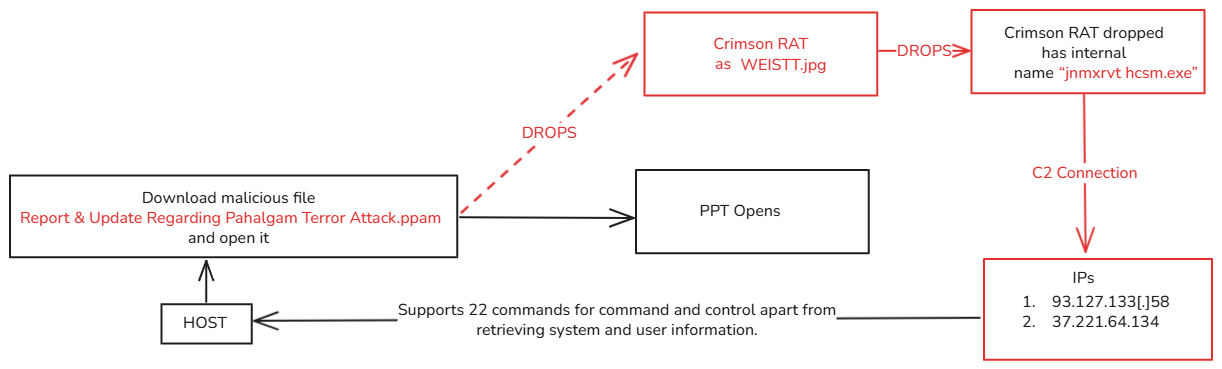}
    \caption{Attack Chain of decoy documents}
    \label{fig:sample_image}
\end{figure}

While the pdfs were being used for phishing, Macro-enabled files such as files with .ppam, .xlam and .xlsb formats embedded malicious macros which were used to deploy malware upon user interaction. One such power point add-on file named \textit{“Report \& Update Regarding Pahalgam Terror Attack.ppam”}, incorporated malicious Visual Basic for Applications (VBA) macros. Upon activation, these macros extracted embedded resources to a newly created, hidden subdirectory within the current user profile, selected payloads based on the host’s Windows version, and presented a decoy document while covertly deploying the Crimson Remote-Access Trojan (RAT)\cite{cyber}. Crimson RAT, a .NET-based malware, enabled post-exploitation capabilities, including keylogging, screen capture, file manipulation, credential theft, and command execution, supporting 22 distinct command-and-control (C2) functions some of which are listed in Table \ref{tab:keycmds}. Its final binary, compiled on April 21, 2025, under the internal name \textit{jnmxrvt hcsm.exe} was dropped as \textit{WEISTT.jpg}, adhered to APT36’s characteristic naming convention with a Program Database (PDB) path of 
\texttt{C:\textbackslash jnmhxrv cstm\textbackslash jnmhxrv cstm\textbackslash obj\textbackslash Debug\textbackslash jnmhxrv cstm.pdb}. A benign hard-coded IP served as a decoy, concealing the actual C2 endpoint at 93.127.133[.]58. 
Additional C2 servers at 167.86.97[.]58:17854, hosted on virtual private servers in Russia, Germany, and Indonesia, enhanced operational obfuscation, as evidenced by telemetry data
Complete attack chain of this malicious decoy document with file format .ppam can be illustrated as shown in Figure \ref{fig:sample_image}
, highlighting each stage from initial delivery to final payload execution.

\begin{table}
\centering
\caption{Key C2 Commands and Their Functions}
\label{tab:keycmds}
\begin{tabular}{|l|l|l|}
\hline
\textbf{S.\,No.} & \textbf{Command} & \textbf{Functionality}\\
\hline
1 & \texttt{procl / getavs} & Get a list of all processes\\ \hline
2 & \texttt{endpo}          & Kill process based on PID\\ \hline
3 & \texttt{cscreen}        & Get screenshot\\ \hline
4 & \texttt{dowf}           & Download a file from C2\\ \hline
5 & \texttt{file}           & Exfiltrate a file to C2\\ \hline
6 & \texttt{info}           & Get machine info (computer name, username, IP, OS, etc.)\\
\hline
\end{tabular}
\end{table}

Operational sophistication was further demonstrated by transitioning from legacy Poseidon loaders to the modular Ares RAT framework, which emerged as the primary vector of compromise. Ares RAT introduced advanced obfuscation techniques, User Account Control (UAC) bypass mechanisms, and obfuscated PowerShell scripts, leveraging Living Off the Land Binaries (LOLBins) to minimize its detectable footprint. It established connections to C2 server at 167.86.97[.]58:17854, facilitating data exfiltration and operational disruption. 

In addition to the distribution of decoy documents, hacktivist operations, coordinated under hashtags such as \#OpIndia and \#OperationSindoor, amplified the campaign’s impact through distributed denial-of-service (DDoS) attacks, website defacements, and data leaks targeting critical entities, including the Ministry of Defence, National Informatics Centre (NIC), Goods and Services Tax Network (GSTN), All India Institute of Medical Sciences (AIIMS), Jio, and Bharat Sanchar Nigam Limited (BSNL).According to the report from 35 hacktivist groups, including seven newly emergent collectives, contributed to the campaign’s psychological and operational effects\cite{seqrite_2}.

%% file: Logging.tex
In the context of sophisticated cyber warfare campaigns, especially those where hack for hire groups and APTs are involved which uses specially crafted malware and Remote Access Trojans or RATs, It is imperative to collect  high-quality endpoint logs, which can also be co-related, for piecing together adversary actions and inspecting ongoing threats.

As seen in the current example as well, a common technique used by many APT groups to gain initial access to an endpoint is spear-phishing . In this approach, they send specially crafted emails with attachments containing embedded macros, scripts, or other malicious code. These attachments are often files associated with Office-related applications, such as Microsoft Word, Microsoft PowerPoint, and similar formats like PDFs, XLAMs, and others. Since such file types are widely used in daily work, our focus was on collecting logs related specifically to these applications.
When a user opens these files, the hidden malicious code automatically triggers a connection to command-and-control (C2) servers or installs malware.

One problem that arises during the detection of these types of attacks is the large volume of data generated across endpoints. Analyzing this data to identify anomalous or malicious behavior, whether from a single endpoint or from a group of endpoints, and ability to correlate log sources, is extremely challenging. It is equivalent to finding a needle in a haystack, as the malicious activity can be buried under a vast amount of normal, routine events.

While there are various frameworks that can be used for collecting activities and logs from endpoints, to circumvent the problem of high volume and correlation of data, we chose to use Osquery\cite{osquery}\cite{osquery_1} tool. 
Osquery is an open-source, community-driven cross-platform instrumentation framework that allows the endpoint's properties and activities to be queried like a relational database. It exposes system state as tables, enabling real-time collection and analysis of activities done on the endpoint. Using SQL queries, investigators can retrieve details related to process activity, file modifications, network connections, and other operating system activities in form of system events, from across the environment. 
Osquery is compatible with Linux, macOS, Windows, and FreeBSD, featuring around 280 virtual tables in total—56 of which are shared across all prominent platforms—and platform-specific coverage of about 155 tables for Linux, 184 for macOS, and 113 for Windows.
Osquery operates in two modes: interactive and daemon. The interactive mode allows manual querying of live system information, while the daemon mode supports scheduled queries and the collection of event-based telemetry. This architecture supports monitoring system behaviour in real time and over time which in turns allows us to correlate logs and use  several heuristics such as parent-child heuristic and unusual port heuristic etc. which helps us in identifying malicious activities that may not be immediately obvious.

Osquery collects logs of several events related to processes, sockets and files. However, all event logs may not be sufficient to get visibility about activities of several malware including the RATs used in this particular campaign as well. A key advantage of Osquery is its flexible, modular design. Using a Thrift-based RPC interface, you can build external extensions to add new virtual tables or connect other data sources. This makes it easy to customize the framework for specific monitoring needs and expand its capabilities beyond the built-in tables.
To enhance our log-capturing capabilities, we integrated a community \cite{ref_url2} extension into the Osquery deployment. This extension allows us to directly collect logs from Windows-based endpoints and present them in a relational table format.
% A common tactic used by many APT groups is spear-phishing to gain initial access to an endpoint. In this approach, they send specially crafted emails with attachments containing embedded macros, scripts, or other malicious code. When a user opens these files, the code automatically triggers a connection to command-and-control (C2) servers or installs malware

This setup allows us to run targeted queries and collect only the logs relevant to our investigation. For example, we can design queries that look for processes spawned by Microsoft Word that initiate outbound network connections, monitor for unusual file writes by PowerPoint processes in system directories, or detect PDF readers that execute scripts or tries to make connections to external servers. By focusing on these specific behaviours, we can filter out irrelevant noise and zero in on potential malicious activity.

% These attachments are often files associated with Office-related applications, such as Microsoft Word, Microsoft PowerPoint, and similar formats like PDFs, XLAMs, and others. Since such file types are widely used in daily work, our focus was on collecting logs related specifically to these applications.

% While there are various frameworks for detecting malicious activity across different malware categories, we chose to use osquery along with an openly available community extension. This setup allows us to run targeted queries and collect only the logs relevant to our investigation. For example, we can design queries that look for processes spawned by Microsoft Word that initiate outbound network connections, monitor for unusual file writes by PowerPoint processes in system directories, or detect PDF readers that execute scripts. By focusing on these specific behaviors, we can filter out irrelevant noise and zero in on potential malicious activity.for easy querying, thereby improving our visibility into important activities such as process execution, network connections, file system changes, and the creation of executable files. It does this using evented tables like win\_process\_events, win\_socket\_events, win\_file\_events, win\_http\_events, win\_ssl\_events, and win\_dns\_events.

By combining and analyzing this collected data — such as process execution events, file changes, or network socket activity — we can trace all actions performed by a malicious binary or its child processes. For example, process execution details from $win\_process\_events$ can be compared with threat intelligence lists containing known malicious file hashes. Similarly, an outbound network connection recorded in $win\_socket\_events$ from a child process of a trusted binary can be checked against threat intelligence data containing IP addresses and domains linked to command-and-control (C2) servers.

A practical analytical sequence could involve spotting an unusual process creation event stemming from a non-standard parent process in $win\_process\_events$, followed by identifying file modifications linked to that process in $win\_file\_events$. If the process hash or any of the contacted remote IP addresses are malicious upon checking with intelligence feeds, the correlation offers high-confidence attribution of malicious activity. Such a workflow not only identifies the existence of known threats but may also uncover suspicious behavioural patterns that suggest new or previously unclassified malware.

By combining these data sources, security teams can build detailed attack stories. When process creation, file changes, and network activity are connected to external intelligence, analysts can turn separate endpoint events into a clear timeline of an intrusion, enabling faster containment and response. This shows that Osquery, when supported by manual threat hunting and threat intelligence, can be used not only to detect threats but also to reconstruct attack chains during advanced investigations.

%% file: Telemetry.tex
To comprehensively analyze the malicious activities, telemetry data was collected using Osquery, combined with custom extensions to enhance detection capabilities. This section elucidates the telemetry data, detailing the structure and significance of the recorded events across three primary tables—win\_process\_events,  win\_file\_events, and win\_socket\_events—to provide insights into the malware’s behavior, persistence mechanisms, and network interactions.

\subsection{Telemetry Collection Methodology}

\begin{figure}[!ht]
    \centering
    \includegraphics[scale=0.4]{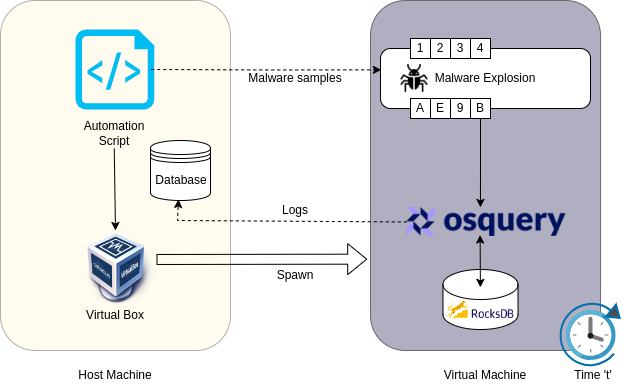}
    \caption{Architecture of the Setup}
    \label{fig:architecture.png}
\end{figure}

To investigate the behaviour of the malware associated with the Operation Sindoor cyber campaign, a controlled experimental environment was established. A virtual machine (VM) instance running Microsoft Windows 11 (computer name: DESKTOP-3DD3GTB), was created using Oracle VirtualBox. To prevent unintended propagation of malware to physical systems, the virtual machine was configured to operate within an isolated network environment, ensuring containment and safeguarding external systems from potential infection.

Within the VM, Osquery was deployed in daemon mode to facilitate continuous, real-time monitoring of system activities. The Osquery configuration incorporated the \textit{--logger\_plugin=filesystem} flag, enabling the logging of detailed telemetry data, including process executions, registry modifications, file operations, and network connections. These logs were stored in a file named osqueryd.results.log, which served as the primary dataset for subsequent analysis. Custom extension was integrated, enhancing the granularity of telemetry collection from the Windows system. This extension allowed for the capture of detailed event data, critical for analyzing the sophisticated behaviours of the Crimson Remote-Access Trojan (RAT) deployed in the campaign.
The extension also helps keep a track of individual event through a unique event ID and individual instance of a process through its GUID.

A malicious sample, named Report \& Update Regarding Pahalgam Terror Attack.pdf, with the SHA-256 hash \[8cbd47119356081e70fc023d3ac78af560651e7932636adeca7bec96b09e0e95,\] was downloaded into the VM. This file, identified as a malicious PowerPoint add-in (*.ppam), was executed to initiate the infection process. Following the opening of the document, a 600-second observation period was enforced to allow the malware to exhibit its behaviours, including process spawning, file operations, and network communications. After this period, the osqueryd.results.log file was securely transferred to a separate system for detailed analysis, ensuring no residual malicious activity affected the analysis environment. The sandbox setup  used to trigger the malware is as shown in Figure \ref{fig:architecture.png}.

To trace the malware’s activities, a forward tracking approach was employed, leveraging the telemetry captured in osqueryd.results.log. This methodology utilized the SHA-256 hash of the initial malicious file to identify associated processes and their behaviors across multiple Osquery tables, specifically win\_process\_events, win\_file\_events, and win\_socket\_events

The telemetry data, identified by the malware hash \[8cbd47119356081e70fc023d3ac78af560651e7932636adeca7bec96b09e0e95,\] was collected, with key events centered around the execution of a malicious PowerPoint add-in file \[8cbd47119356081e70fc023d3ac78af560651e7932636adeca7bec96b09e0e95.ppam.\] The following subsections describe the structure and implications of each telemetry table.

\subsection{Process Events(win\_process\_events)}

The win\_process\_events table, records process creation and termination events, providing insights into the execution chain of the malware. The column descriptions for win\_process\_events are provided in Table \ref{tab:win_process_events}

\begin{table}
\centering
\caption{Columns of \texttt{win\_process\_events}}
\label{tab:win_process_events}
\begin{tabular}{|l|l|l|}
\hline
\textbf{S.\,No.} & \textbf{Column}               & \textbf{Description}\\
\hline
1  & \texttt{time}                  & Event timestamp in UNIX epoch seconds\\ \hline
2  & \texttt{action}                & Type of process event (e.g., \texttt{PROC\_CREATE}, \texttt{PROC\_EXIT})\\ \hline
3  & \texttt{pid}                   & Process identifier of the event’s target process\\ \hline
4  & \texttt{parent\_pid}           & Process identifier of the parent process\\ \hline
5  & \texttt{path}                  & Full filesystem path to the executable\\ \hline
6  & \texttt{cmdline}               & Complete command-line string used to launch the process\\ \hline
7  & \texttt{user}                  & Security principal under which the process ran (DOMAIN\textbackslash user)\\ \hline
8  & \texttt{process\_guid}         & Globally unique identifier for this process instance\\ \hline
9  & \texttt{parent\_process\_guid} & GUID of the parent process instance\\
\hline
\end{tabular}
\end{table}

\begin{figure}[!ht]
    \centering
    \includegraphics[width=1.0\textwidth]{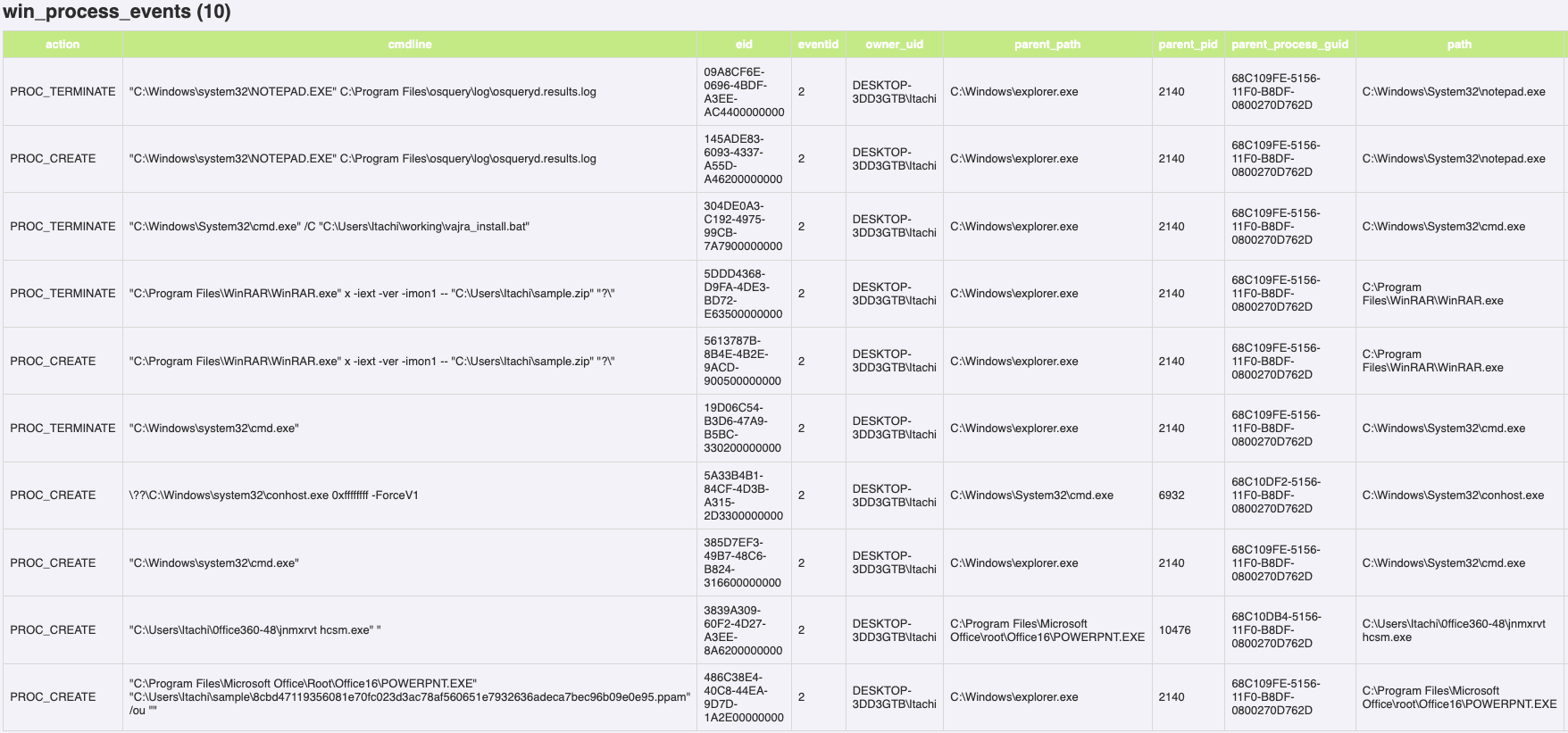}
    \caption{Process Related Activities}
    \label{fig:WinProcessEvents.png}
\end{figure}

Notable events include two pivotal malware-related events were logged, delineating the initial infection and payload deployment stages of the Crimson RAT. At 14:33:44 UTC, a PROC\_CREATE event (event ID 486C38E4-40C8-44EA-9D7D-1A2E00000000) recorded the execution of POWERPNT.EXE (PID 10476, process GUID 68C10DB4-5156-11F0-B8DF-0800270D762D) with the command line \[C:\backslash Program Files\backslash Microsoft Office\backslash Root\backslash Office16\backslash POWERPNT.EXE\]

\[C:\backslash Users\backslash Itachi\backslash sample\backslash\] \[8cbd47119356081e70fc023d3ac78af560651e7932636adeca7bec96b09e0e95.ppam. /ou\]
This process, initiated by explorer.exe (PID 2140, parent process GUID 68C109FE-5156-11F0-B8DF-0800270D762D) under the user account DESKTOP-3DD3GTB\textbackslash{Itachi}, marks the infection’s onset. The .ppam file, a macro-enabled PowerPoint add-in with a matching SHA-256 hash, likely contains malicious Visual Basic for Applications (VBA) macros or scripts that trigger subsequent malicious actions. The /ou flag suggests the file was opened in a specific mode, potentially bypassing user prompts to enable macros automatically. Five seconds later, at 14:33:49 UTC, a second PROC\_CREATE event (event ID 3839A309-60F2-4D27-A3EE-8A6200000000) recorded POWERPNT.EXE spawning jnmxrvt hcsm.exe (PID 2192, process GUID 68C10DCA-5156-11F0-B8DF-0800270D762D) in C:\textbackslash{Users}\textbackslash{Itachi}\textbackslash{0ffice360-48}\textbackslash with the command line "C:\textbackslash{Users}\textbackslash{Itachi}\textbackslash{0ffice360-48}\textbackslash{jnmxrvt hcsm.exe}" as can be seen in Figure \ref{fig:WinProcessEvents.png}. This child process, executed under the same user account, represents the Crimson RAT, responsible for establishing persistence and initiating C2 communications. The creation of jnmxrvt hcsm.exe signifies a critical transition in the infection chain, where the initial macro-driven infection evolves into a fully functional RAT. These events, executed within a tight five-second window, highlight APT36’s sophisticated use of legitimate Microsoft Office processes to mask malicious intent, leveraging the trusted POWERPNT.EXE to deploy a persistent threat.

\subsection{File Events(win\_file\_events)}
The win\_file\_events table, logs file creation, writing, deletion, and renaming activities, crucial for tracking the malware’s filesystem interactions. The column descriptions for win\_file\_events are provided in Table \ref{tab:win_file_events}

\begin{table}
\centering
\caption{Columns of \texttt{win\_file\_events}}
\label{tab:win_file_events}
\begin{tabular}{|l|l|l|}
\hline
\textbf{S.\,No.} & \textbf{Column}        & \textbf{Description}\\
\hline
1  & \texttt{time}          & Event timestamp in UNIX epoch seconds\\ \hline
2  & \texttt{action}        & File operation type (e.g., \texttt{FILE\_CREATE}, \texttt{FILE\_DELETE})\\ \hline
3  & \texttt{eid}           & Unique event identifier (UUID) for deduplication\\ \hline
4  & \texttt{target\_path}  & Absolute path of the file being accessed\\ \hline
5  & \texttt{md5}           & MD5 checksum of the file at event time\\ \hline
6  & \texttt{sha256}        & SHA-256 checksum of the file at event time\\ \hline
7  & \texttt{pid}           & Process identifier that performed the file action\\ \hline
8  & \texttt{process\_guid} & GUID of the process instance performing the action\\ \hline
9  & \texttt{process\_name} & Name (and path) of the executable process\\
\hline
\end{tabular}
\end{table}

\begin{figure}[!ht]
    \centering
    \includegraphics[scale=0.4]{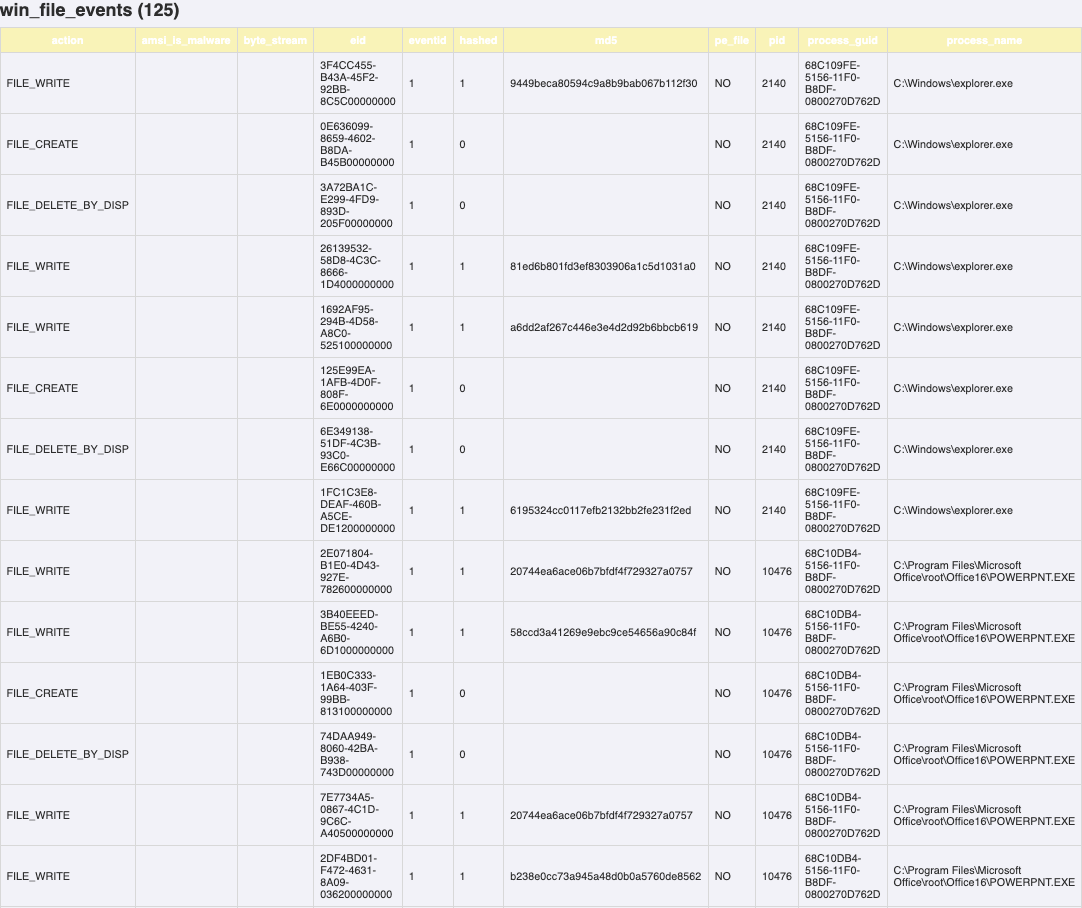}
    \caption{File Related Activities}
    \label{fig:WinFileEvents.png}
\end{figure}

The fields captured include action, which denotes operations like FILE\_CREATE (creating a new file), FILE\_WRITE (writing data to a file), or FILE\_RENAME (renaming a file); process\_name (C:\textbackslash Program Files\textbackslash Microsoft Office\textbackslash Root\textbackslash Office16\textbackslash POWERPNT.EXE), identifying the legitimate Microsoft PowerPoint executable exploited by the malware; target\_path, specifying the file’s location; md5 and sha256, providing cryptographic hashes for file verification; pid (10476) and process\_guid (68C10DB4-5156-11F0-B8DF-0800270D762D), linking events to the malicious process; time and utc\_time, timestamping events (e.g., 1752503624 for 14:33:44 UTC); and eid, a unique event identifier. 
A major event occurred at 14:33:49 UTC, when POWERPNT.EXE executed a FILE\_CREATE action to create WEISTE.jpg in \textit{C:\textbackslash Users\textbackslash Itachi\textbackslash 0ffice360-48\textbackslash}
 This file, disguised as a benign image, likely served as a dropper containing the encoded RAT executable, a tactic used to bypass initial security checks by masquerading as a non-executable file. The non-standard directory 0ffice360-48 (with a deliberate misspelling of “Office”) suggests an attempt to evade detection by avoiding typical Office paths like \texttt{C:\textbackslash Program Files\textbackslash Microsoft Office\textbackslash}. 
 Immediately following, at 14:33:49 UTC, a FILE\_RENAME event renamed WEISTE.jpg to jnmxrvt hcsm.exe as can be seen in Figure \ref{fig:WinFileEvents.png}, revealing its true purpose as the Crimson RAT executable (later executed as PID 2192, process GUID 68C10DCA-5156-11F0-B8DF-0800270D762D). This rename is a critical step, transitioning the infection from a disguised dropper to an executable capable of persistence and C2 communications. Concurrently, at around 14:33:48 UTC, POWERPNT.EXE created and wrote files including vbaProject.bin and oleObject1.bin through oleObject5.bin in "C:\textbackslash Users\textbackslash Itachi\textbackslash 0ffice360-48\textbackslash". The vbaProject.bin
file likely contains malicious VBA macros, which are scripts that execute automatically to unpack or download payloads, while the oleObject*.bin files suggest embedded objects (e.g., scripts or executables) crucial for staging the RAT. These files, created in the same non-standard directory, further indicate stealth tactics. At 14:33:45 UTC, a FILE\_CREATE event generated 
\[8cbd47119356081e70fc023d3ac78af560651e7932636adeca7bec96b09e0e95.ppam,\] followed by FILE\_WRITE event at 14:33:50 UTC matching the malicious file’s SHA-256 hash, confirming its role as the primary infection vector—a macro-enabled PowerPoint add-in that triggers malicious scripts upon opening. Additionally, temporary files were created and written in 
\[C:\backslash Users\textbackslash Itachi\backslash AppData\backslash Local\backslash Temp\backslash,\]
a directory commonly used by malware for transient payload staging to minimize forensic evidence.

\subsection{Socket Events (win\_socket\_events)} The win\_socket\_events table, records network connections, essential for identifying command-and-control (C2) communications. 
The column descriptions for win\_socket\_events are provided in Table \ref{tab:win_socket_events}.

\begin{table}
\centering
\caption{Columns of \texttt{win\_socket\_events}}
\label{tab:win_socket_events}
\begin{tabular}{|l|l|l|}
\hline
\textbf{S.\,No.} & \textbf{Column}           & \textbf{Description}\\
\hline
1  & \texttt{time}              & Event timestamp in UNIX epoch seconds\\ \hline
2  & \texttt{action}            & Socket event type (e.g., \texttt{connect}, \texttt{accept}, \texttt{close})\\ \hline
3  & \texttt{pid}               & Process identifier owning the socket\\ \hline
4  & \texttt{process\_guid}     & GUID of the owning process instance\\ \hline
5  & \texttt{process\_name}     & Name (and path) of the process executable\\ \hline
6  & \texttt{family}            & Address family (e.g., \texttt{AF\_INET} for IPv4)\\ \hline
7  & \texttt{local\_address}    & Local IP address for the socket endpoint\\ \hline
8  & \texttt{local\_port}       & Local port number used by the socket\\ \hline
9  & \texttt{remote\_address}   & Remote IP address the socket connected to\\ \hline
10 & \texttt{remote\_port}      & Remote port number on the peer endpoint\\
\hline
\end{tabular}
\end{table}

A detailed record of network connection attempts reveals the command-and-control (C2) communication patterns used by the Crimson RAT during the Operation Sindoor campaign.
The fields include action (SOCKET\_CONNECT), process\_name, like
\[C:\backslash Program Files\backslash Microsoft Office\backslash Root\backslash Office16\backslash POWERPNT.EXE\]
 or 
 \[C:\backslash Users\backslash Itachi\backslash 0ffice360-48\backslash jnmxrvt hcsm.exe\]
, remote\_address, remote\_port, local\_address (10.0.2.15), local\_port, pid, process\_guid, time, utc\_time, family (AF\_INET), protocol (TCP, 6), event\_type (SOCKET), and eid.

The process POWERPNT.EXE (PID 10476, GUID 68C10DB4-5156-11F0-B8DF-0800270D762D) initiated three SOCKET\_CONNECT events to 37.221.64.134:443, a phishing link resolving to \url{jkpolice[.]gov[.]in[.]kashmirattack[.]exposed}, at 14:34:40 UTC (event ID 753FB770-923A-42B2-9493-157103000000, port 55626), 14:35:18 UTC (timestamp 1752503718, event ID 4DE4180B-FCFF-48E9-810E-2F3F03000000, port 55643), and 14:35:33 UTC (timestamp 1752503733) and event ID 6130D1D4-DC08-41DB-9FB8-877303000000, port 55647). Occurring 56–109 seconds after the malicious .ppam execution at 14:33:44 UTC, these connections on port 443 (HTTPS) suggest potential phishing activity, leveraging the legitimate appearance of Office traffic to deceive users or deliver malicious payloads. 

\begin{figure}[!ht]
    \centering
    \includegraphics[scale=0.4]{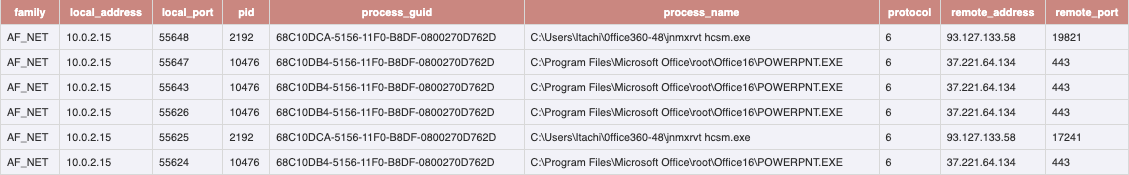}
    \caption{Socket Related Activities}
    \label{fig:WinSocketEvents.png}
\end{figure}

The domain’s deceptive naming mimics an official Jammu and Kashmir Police site, a tactic consistent with APT36’s social engineering strategies. Concurrently, jnmxrvt hcsm.exe (PID 2192, GUID 68C10DCA-5156-11F0-B8DF-0800270D762D) initiated two SOCKET\_CONNECT events to 93.127.133.58:19821, a confirmed C2 IP, at 14:34:35 UTC (timestamp 1752503675) and 14:35:38 UTC (timestamp 1752503738), both with event ID 65D9BA08-9728-46A1-BCBA-CE0003000000 and local port 55648 and 55625, 51–114 seconds after the RAT’s execution at 14:33:49 UTC as can be seen in Figure \ref{fig:WinSocketEvents.png}. 
These TCP connections, using a non-standard port, indicate C2 activity for command retrieval or data exfiltration, aligning with the Crimson RAT’s role in maintaining attacker control. The identical eid suggests a logging artifact or repeated connection attempts to ensure C2 server connectivity, possibly to bypass network restrictions. 

The rapid onset of network activity post-infection highlights APT36’s sophisticated tactics, using phishing to initiate compromise and C2 for persistence and exfiltration. Distinguishing POWERPNT.EXE’s phishing traffic from jnmxrvt hcsm.exe’s C2 communications requires deep packet inspection and threat intelligence to confirm the malicious nature of 37.221.64.134 and 93.127.133.58. Strong monitoring systems, such as Endpoint Detection and Response (EDR) tools \cite{vajra2} or Extended Detection and Response (XDR) tools, are essential for detecting these types of activities and providing response capabilities, especially when monitoring a large number of endpoints across an organization.

\begin{table}[h]
\centering
\caption{Structure of Osquery SQL Query for Operation Sindoor}
\label{tab:osquery_structure}
\begin{tabular}{|p{2.5cm}|p{5.5cm}|}
\hline
\textbf{Component} & \textbf{Description} \\
\hline
\texttt{SELECT} & Retrieves fields (\texttt{table\_name}, \texttt{action}, \texttt{process\_name}, \texttt{target\_path}/\texttt{target}, \texttt{md5}, \texttt{sha256}, \texttt{utc\_time}). \\
\texttt{FROM} & Queries \texttt{win\_process\_events} ,\texttt{win\_file\_events} and \texttt{win\_socket\_events}. \\
\texttt{WHERE} & Filters file events (\texttt{POWERPNT.exe}, \texttt{jnmxrvt hcsm.exe}, hashes, \texttt{WEISTE.jpg} rename) and network events (suspicious IPs/ports, excluding legitimate traffic). \\
\texttt{UNION} & Combines file and network results, mimicking Sigma’s OR condition. \\
\hline
\end{tabular}
\end{table}

The telemetry data illustrates a multi-stage attack chain initiated by user interaction with a malicious PowerPoint add-in, leading to the deployment of Crimson RAT. The process events confirm the execution of jnmxrvt hcsm.exe from a dynamically created directory 
\[C:\backslash Users\backslash Itachi\backslash 0ffice360-48\backslash,\] while file events reveal the creation of malicious payloads disguised as Office components, and socket events confirm C2 communications, highlighting the malware’s espionage capabilities. Osquery’s granular logging, enhanced by custom extensions, enabled the detection of these activities by correlating process, file, registry, and network events, providing a comprehensive view of the attack lifecycle.

The rapid deployment of the malware, compiled on April 21, 2025, and executed in July, underscores APT36’s agility in exploiting geopolitical events like the Pahalgam attack.

%The structure of this query is based on Table \ref{tab:osquery_structure}
%and the actual query can be founding in Listing \ref{lst:osquery}

Based on this telemetry we will create a query for Osquery which can be used as a rule to detect any attacks related to this campaign. The Osquery SQL query replicates the Sigma rule’s logic, querying \texttt{win\_file\_events} and \texttt{win\_socket\_events} to detect file operations and network connections associated with the Crimson RAT. The query’s components are summarized in Table~\ref{tab:osquery_structure}, with the full query presented in Listing~\ref{lst:osquery}.

\begin{lstlisting}[caption={Osquery SQL Query for Detecting Crimson RAT Artifacts},label={lst:osquery}]
-- Files
SELECT 
  'win_file_events' AS table_name,
  action,
  process_name,
  target_path,
  md5,
  utc_time
FROM win_file_events
WHERE
  (
    -- known hashes
    process_name ILIKE '%\\msedge.exe'
    AND action IN ('FILE_RENAME', 'FILE_CREATE')
    AND md5 IN (
      'd946e3e94fec670f9e47aca186ecaabe',
      'e18c4172329c32d8394ba0658d5212c2',
      '2fde001f4c17c8613480091fa48b55a0',
      'c1f4c9f969f955dec2465317b526b600',
      '026e8e7acb2f2a156f8afff64fd54066',
      'fb64c22d37c502bde55b19688d40c803',
      '70b8040730c62e4a52a904251fa74029',
      '3efec6ffcbfe79f71f5410eb46f1c19e',
      'b03211f6feccd3a62273368b52f6079d'
    )
  )
  OR (
    -- Suspicious jnmxrvt hcsm.exe file ops OR WEISTE.jpg drops
    process_name ILIKE '%\\jnmxrvt hcsm.exe'
    AND action IN ('FILE_RENAME', 'FILE_CREATE')
    AND (
      target_path ILIKE '%jnmxrvt hcsm.exe'
      OR target_path ILIKE '%WEISTE.jpg%'
    )
  )
  OR (
    -- Any rename to the suspicious name
    action = 'FILE_RENAME'
    AND target_path ILIKE '%jnmxrvt hcsm.exe'
  )

UNION ALL

-- Sockets
SELECT 
  'win_socket_events' AS table_name,
  action,
  process_name,
  (remote_address || ':' || CAST(remote_port AS TEXT)) AS target_path,
  NULL AS md5,
  utc_time
FROM win_socket_events
WHERE
  (
    process_name ILIKE '%\\POWERPNT.EXE'
    OR process_name ILIKE '%\\jnmxrvt hcsm.exe'
  )
  AND remote_address IN (
    '93.127.133.58',
    '104.129.27.14',
    '37.221.64.134',
    '78.40.143.189',
    '45.141.58.224',
    '45.141.59.167',
    '45.141.58.33',
    '78.40.143.98',
    '84.54.51.12',
    '176.65.143.215',
    '45.141.59.72',
    '192.64.118.76'
  )
  AND remote_port IN (1097, 17241, 19821, 21817, 23221, 
  27425, 8108, 16197, 19867, 28784, 30123)
  AND NOT (remote_address = '96.17.168.104' AND remote_port = 443);

\end{lstlisting}

%% file: Conclusion.tex
We studied the malware campaign launched by Pakistan's APT groups and hacktivist against India during the Operation Sindoor. Following the clues for the behavioural aspected of the cyber campaign released by the Indian threat intelligence agencies, we developed a details of the event logs needs to be monitored and analysed to detect the malware. We ran the malware in sandbox environment and build a telemetry using Osquery with custom extension. We developed a SQL based detection rules to identify presence of the malware and identify active exploitation it performed. It also provided an effective means to block the malware from any potential damages. 

The analysis revealed several other information. The compilation date of malware revealed that it was specifically crafted few days before the Pahalgam attacks, indicating that the adversaries have prior knowledge of the terror attacks and launched the cyber attacks to steal information and build false public perception. This coordination between the terrorist activities and APT groups, further strengthen the claims of Indian intelligences agencies that Pahalgam terrorist attacks were planned and controlled from Pakistani soil.

%% file: main.bbl
\begin{thebibliography}{10}
\providecommand{\url}[1]{\texttt{#1}}
\providecommand{\urlprefix}{URL }
\providecommand{\doi}[1]{https://doi.org/#1}

\bibitem{Russia-Ukraine}
Mueller, G.B., Jensen, B., Valeriano, B., Maness, R.C., Macias, J.M.: Cyber operations during the russo-ukrainian war. \url{https://www.csis.org/analysis/cyber-operations-during-russo-ukrainian-war} (2025)

\bibitem{Iran-Israel}
Cyber threats linked to iran-israel conflict. \url{https://reliaquest.com/blog/cyber-threats-linked-to-iran-israel-conflict/} (June 23 2025)

\bibitem{India-Pakistan}
Capsnetdroff: Cyber warfare: Dual operational fronts in contemporary india-pakistan conflicts. \url{https://capsindia.org/cyber-warfare-dual-operational-fronts-in-contemporary-india-pakistan-conflicts/} (June 5 2025)

\bibitem{Thailand-Cambodia}
Cross-border cyberattacks surge as thailand--cambodia tensions escalate. \url{https://cyberdefensewire.com/cross-border-cyberattacks-surge-as-thailand-cambodia-tensions-escalate/} (July 29 2025)

\bibitem{PIB}
Operation sindoor: India’s strategic clarity and calculated force. \url{https://www.pib.gov.in/PressNoteDetails.aspx?NoteId=154448&ModuleId=34} (May 12 2025)

\bibitem{mitre_attack}
Mitre att\&ck {\textregistered}. \url{https://attack.mitre.org/}

\bibitem{seqrite_1}
Kanjilal, R.: Advisory: Pahalgam attack themed decoys used by apt36 to target the indian government. \url{https://www.seqrite.com/blog/advisory-pahalgam-attack-themed-decoys-used-by-apt36-to-target-the-indian-government/} (June 12 2025)

\bibitem{seqrite_2}
Seqrite: Operation sindoor -- anatomy of a digital siege. \url{https://www.seqrite.com/blog/operation-sindoor-anatomy-of-a-digital-siege/} (May 27 2025)

\bibitem{osquery}
Rapid: Introduction to osquery for threat detection and dfir. \url{https://www.rapid7.com/blog/post/2016/05/09/introduction-to-osquery-for-threat-detection-dfir/} (May 9 2016), rapid7 Blog

\bibitem{osquery_1}
Sql powered operating system instrumentation, monitoring, and analytics. \url{https://github.com/osquery/osquery}

\bibitem{cyber_Survey}
Sajid, A., Razzaq, H., Malik, R., Khan, A.A., Iqbal, M.S., Farhan, S.: Survey paper on cyber warfare. IETI Transactions on Data Analysis and Forecasting (iTDAF)  \textbf{2}(3),  pp. 27–37 (Oct 2024). \doi{10.3991/itdaf.v2i3.51025}, \url{https://online-journals.org/index.php/iTDAF/article/view/51025}

\bibitem{cyber_scanning}
Bou-Harb, E., Debbabi, M., Assi, C.: Cyber scanning: A comprehensive survey. IEEE Communications Surveys \& Tutorials  \textbf{16}(3),  1496--1519 (2014). \doi{10.1109/SURV.2013.102913.00020}

\bibitem{khandait2021}
Khandait, P., Tiwari, N., Hubballi, N.: Who is trying to compromise your ssh server? an analysis of authentication logs and detection of bruteforce attacks. In: Adjunct Proceedings of the 2021 International Conference on Distributed Computing and Networking (ICDCN '21). pp. 127--132. Association for Computing Machinery, New York, NY, USA (2021). \doi{10.1145/3427477.3429772}

\bibitem{prasad2024}
Prasad, M.D., Ali, M.W., Sindusha, M.S.N.V.R.S., Jahnavi, N., Devi, S.A.: Enabling cybersecurity defenses: Advanced endpoint detection, data breach identification and anomaly resolution. In: 2024 8th International Conference on Inventive Systems and Control (ICISC). pp. 461--468 (2024). \doi{10.1109/ICISC62624.2024.00084}

\bibitem{vajra1}
Bakshi, A., Sawant, T., Thakare, P., Dandawala, A., Hanawal, M., Kabra, A.: Improving threat detection capabilities in windows endpoints with osquery. In: 2023 15th International Conference on Communication Systems \& Networks (COMSNETS). pp. 432--435. Bangalore, India (2023). \doi{10.1109/COMSNETS56262.2023.10041379}

\bibitem{vajra2}
Agarwal, S., Sable, A., Sawant, D., Kahalekar, S., Hanawal, M.: Threat detection and response in linux endpoints. In: 2022 14th International Conference on Communication Systems \& Networks (COMSNETS). pp. 447--449. Bangalore, India (2022). \doi{10.1109/COMSNETS53615.2022.9668567}

\bibitem{mosoiu2020}
Mo{\c{s}}oiu, O., B{\u{a}}l{\u{a}}ceanu, I., Mihai, E.: Cyber terrorism and the effects of the russian attacks on democratic states in east europe. Scientific Journal of Silesian University of Technology. Series Transport  \textbf{106},  131--139 (2020). \doi{10.20858/sjsutst.2020.106.11}

\bibitem{aggreyanalysing}
Aggrey, R., Adjei, B., Afoduo, K., Dsane, N., Cudjoe, A., Ababio, M.: Analysing recent apt incidents: case studies and lessons learned

\bibitem{obrien2022}
O'Brien, N.: Assessing the importance of modern security tools and frameworks to help detect and defend against cozy bear (2022), internship Report, National College of Ireland

\bibitem{sarowa2022}
Sarowa, S., Bhanot, B., Kumar, V.: Analysis of cyber attacks and cyber incident patterns over apcert member countries. In: Proceedings of the 2022 4th International Conference on Artificial Intelligence and Speech Technology (AIST). IEEE (2022). \doi{10.1109/AIST55798.2022.10064961}

\bibitem{stango2009}
Stango, A., Prasad, N., Kyriazanos, D.: A threat analysis methodology for security evaluation and enhancement planning. In: Proceedings of the 2009 Third International Conference on Emerging Security Information, Systems and Technologies (SECURWARE 2009). pp. 262--267 (2009). \doi{10.1109/SECURWARE.2009.47}

\bibitem{cyber_power}
Du, D., Zhu, M., Li, X., Fei, M., Bu, S., Wu, L., Li, K.: A review on cybersecurity analysis, attack detection, and attack defense methods in cyber-physical power systems. Journal of Modern Power Systems and Clean Energy  \textbf{11}(3),  727--743 (2023). \doi{10.35833/MPCE.2021.000604}

\bibitem{cyber_iiot}
Alnajim, A.M., Habib, S., Islam, M., Thwin, S.M., Alotaibi, F.: A comprehensive survey of cybersecurity threats, attacks, and effective countermeasures in industrial internet of things. Technologies  \textbf{11}(6) (2023). \doi{10.3390/technologies11060161}, \url{https://www.mdpi.com/2227-7080/11/6/161}

\bibitem{APT36}
Apt profile: Transparent tribe aka apt36 - cyfirma. \url{https://www.cyfirma.com/research/apt-profile-transparent-tribe-aka-apt36/} (May 15 2025)

\bibitem{cyber}
Team, C.R.: Cyber attacks rise as tension mounts across india pakistan border post terrorist attack. \url{https://www.cyberproof.com/blog/cyber-attacks-rise-as-tension-mounts-across-india-pakistan-border-post-terrorist-attack/} (May 30 2025)

\bibitem{ref_url2}
Extension to osquery windows that enhances it with real-time telemetry, log monitoring and other endpoint data collection. \url{https://github.com/hecg119//eclecticiq-osq-ext-bin}

\end{thebibliography}
